\documentstyle[12pt]{article}

\begin{document}

\def\be{\begin{equation}}
\def\ee{\end{equation}}
\def\tr{{\rm tr}}
\def\nn{\nonumber}
\def\jm{j^{\mu}}
\def\fr{{<\phi^{2}>}_{R}}
\def\g5{{\gamma}_{5}}
\def\sm{$ {\rm T}_{\rm SM} $}
\def\bea{\begin{eqnarray}}
\def\eea{\end{eqnarray}}
\def\ba{\begin{array}}
\def\ea{\end{array}}
\def\pa{{\partial}}
\def\mt{{\rm m}_{t}}
\newcommand{\etal}{\mbox{\it et al.}}
\newcommand{\prdj}[1]{{ \it Phys.~Rev.}~{\bf D{#1}}}
\newcommand{\prlj}[1]{{ \it Phys.~Rev.~Lett.}~{\bf {#1}}}
\newcommand{\plbj}[1]{{ \it Phys.~Lett.}~{\bf {#1B}}}
\newcommand{\npbj}[1]{{ \it Nucl.~Phys.}~{\bf B{#1}}}


\begin{titlepage}
\begin{flushright}   BONN-TH-99-08  \end{flushright}
\begin{flushright}   IFT-9/99  \end{flushright}
\begin{flushright}   hep-th/9905139  \end{flushright}
\vskip .5cm
\centerline{\LARGE{\bf {Supersymmetry Breakdown}}}
\vskip .5cm
\centerline{\LARGE{\bf {at Distant Branes:}}}
\vskip .5cm
\centerline{\LARGE{\bf {The Super--Higgs Mechanism}}}
\vskip 1.5cm
\centerline{\bf Krzysztof A. Meissner${}^*$,
Hans Peter Nilles${}^{\#}$ and Marek Olechowski${}^{*}$}

\vskip .5cm
\centerline{\em ${}^{*}$ Institute of Theoretical Physics}
\centerline{\em ul. Ho\.za 69}
\centerline{\em 00-681 Warszawa, Poland}

\vskip .5cm
\centerline{\em ${}^{\#}$ Physikalisches Institut, Universit\"at Bonn}
\centerline{\em Nussallee 12}
\centerline{\em D-53115 Bonn, Germany}

\vskip 1.5cm
\centerline{\bf {Abstract}}
\indent

A compactification of 11-dimensional supergravity with two (or more) 
walls is considered. The whole tower of massive Kaluza-Klein modes along 
the fifth dimension is taken into account. With the sources on the walls, 
an explicit composition in terms of Kaluza-Klein modes of massless 
gravitino (in the supersymmetry preserving case) and massive gravitino 
(in the supersymmetry breaking case) is obtained. The super--Higgs effect 
is discussed in detail.

\vfill
\noindent
May 1999

\end{titlepage}


\section{Introduction}

The M--theoretic extension of the heterotic $E_8\times E_8$ string
leads to a geometric picture of two walls (branes) at the
ends of a finite 11-dimensional interval \cite{HoravaWitten}. 
While the supergravity
multiplet can penetrate in the $d=11$ bulk, the two $E_8$
gauge multiplets are confined to the two walls, respectively.
When further six of the dimensions are compactified one can 
construct models with gauge fields living on the $d=4$ walls
while gravity could, in addition, extend to the higher
dimensional interval \cite{Witten}. This could then be viewed
as the M-theoretic generalization of hidden sector supergravity
models \cite{PR}. These models typically contain two sectors,
an observable sector that contains the usual fields like
quarks, leptons and gauge bosons as well as their supersymmetric
partners and a hidden sector, coupled to the observable sector
via interactions of gravitational strength. Here the two sectors
can now be identified with the gauge systems that live on the two 
separated walls. Such a picture is common in modern string-brane
theories. The role of the walls is in general played by
higher dimensional p-branes that support gauge groups, while
gravitational interactions can communicate between spatially
seperated branes. In type I theories we have e.g. D-branes
with gauge bosons originating from open string whose ends are
confined to the (stack of coincident) D-branes, while closed
strings (and thus gravitational interactions) can live in the
bulk.

The hidden sector of the above mentioned supergravity models
was supposed to be responsible for the breakdown of 
supersymmetry \cite{HPN}. This breakdown of supersymmetry
was  transmitted to the observable sector via
gravitational interactions. If the breakdown originated through
the vacuum expectation value of an auxiliary field of
size $F= M_S^2$, the value of the gravitino mass is given
by  
\be
   m_{3/2} \sim \frac{M_S^2}{M_{\rm Planck}},
\label{eq:gravitinomass}
\ee
where the Planck mass represents the suppression due to
the gravitational interactions, and the size of the soft
supersymmetry breaking terms in the observable sector
was given by the gravitino mass.

In the modern picture one would now assume that supersymmetry
is broken at a hidden wall \cite{Horava} and the transmission
of that breakdown to the observable wall is mediated via
bulk fields. The size of supersymmetry breakdown in the
observable sector would be suppressed for widely seperated
walls.  Naively one might have assumed that the new picture
would lead to a value of the gravitino mass even more suppressed
than in the classical case  (\ref{eq:gravitinomass}). A closer
inspection, however,  shows a similar suppression \cite{NOY1,NOY2}
\begin{equation}
   m_{3/2} \sim {{M_S^2}\over{R M_{\rm D}^2}}
\sim {{M_S^2}\over{M_{\rm Planck}}},
\label{eq:gravitinomass2}
\end{equation}
once the distance $R$ between the branes and the the higher
dimensional Planck mass $M_{\rm D}$
are adjusted to fit the
correct value of the $d=4$ Planck mass ${M_{\rm Planck}}$.

The estimate of the gravitino mass in (\ref{eq:gravitinomass2})
was obtained \cite{NOY1,Lalak,Lukas}
using a the simplified approximation according
to which the higher dimensional bulk fields were integrated out
via an averaging proceedure\footnote{A corresponding analysis in global
supersymmetry has been performed in ref. \cite{Mirabelli}.
Related work in the supergravity case has been given in \cite{Ellisetc}.}. 
In this picture, the goldstino
mode was represented by the lowest Kaluza--Klein $\Psi_0$ mode of a
higher dimensional field $\Psi$. In the super--Higgs mechanism this
mode supplies the additional degrees of freedom to render the
gravitino massive. Qualitatively this simplified approximation
does give a consistent picture, but there remain some open
questions and potential problems when one looks into details of
the super--Higgs mechanism. 
In this paper we would like to point out these
potential problems and show how they can be resolved. The open
questions will be presented in the following section. In section 3
we shall then discuss the gravitino in the case of unbroken
supersymmetry in full generality. 
Broken supersymmetry and the super--Higgs mechanism
will be analysed in section 4. In the following section we shall
discuss the consequences of our analysis. This will include
a discussion of the possible nature of the goldstino (is it a bulk
or a wall field), the relation to the Scherk-Schwarz mechanism 
\cite{ScherkSchwarz}
in that context \cite{Antoniadis} and an upper limit for
the gravitino mass in the present picture. We shall argue that
a meaningful realization of the super--Higgs mechanism seems to 
require some modes in the bulk other than the graviton and the 
gravitino. Finally we shall comment on the phenomenological
consequences of this findings, including 
a discussion of the nature of the soft
breaking terms on both walls.

\section{Some open questions and puzzles}

Specifically we want to address the following two questions:

\begin{itemize}

\item[(i)] the nature of the massless gravitino in the presence
of several $F-$terms\footnote{In this paper we generically use
the notation $F-$term for the source of supersymmetry breakdown.
Depending on the specific situation this could represent a
$D-$term or a gaugino condensate as well.}
 on different walls that cancel and lead to
unbroken supersymmetry

\item[(ii)] the identification of the goldstino in the case of
broken supersymmetry.

\end{itemize}

The first question (i) arises because of a particular nonlocal effect
of supersymmetry breakdown first observed by Ho\v{r}ava \cite{Horava}.
A given source of supersymmetry breakdown (parametrized by a
vacuum expectation value (vev) of an auxiliary field $F$) on one wall
could be compensated by a similar but opposite value $(-F)$ on 
another (separated) wall. Any calculation and approximation of
the system thus has to reproduce this behaviour. The
previously mentioned averaging proceedure over the bulk distance
does this in a trivial way, leading to unbroken 
supersymmetry as expected.
A detailed inspection of the gravitino, however, reveals a problem.
If we start with the situation $F=0$ it is easy to define the
massless gravitino $\Psi_0$ in the $d=4$ theory. Switching on a nontrivial
$F$ on one brane and $(-F)$ on the other still should give a
massless gravitino, but $\Psi_0$ turns out to be no longer a mass eigenstate.
The resolution of this problem and the correct identification of the
gravitino will be given in section 3. It is a particular
combination of the possible gravitini that appear when one, for
example, reduces a 5-dimensional theory to a theory in $d=4$ on a
finite $d=5$ interval. The theory on a $d=5$ circle would lead to
$N=2$ supersymmetry in $d=4$ and two massless gravitini (zero modes on
the circle). The
$Z_2$ projection on the interval removes one of the gravitini
and is $N=1$ supersymmetric. A nonvanishing vev of $F$ now interferes
with the boundary conditions and the massless gravitino will be
a linear combination of the zero mode and all the excited KK modes whose
coefficients will depend on $F$ (assuming, of course, unbroken
supersymmetry due to a compensating vev $-F$ on another wall).

The second question (ii) deals with the nature of the
goldstino (i.e. the longitudinal components of the gravitino)
in the case of broken supersymmetry. Remember that the
simplified averaging proceedure leads to a goldstino that
corresponds to the lowest Kaluza--Klein mode $\Psi_0$
of a higher-dimensional bulk field $\Psi$. Inspecting the
gravitino mass matrix in this case reveals the fact that this
field $\Psi_0$ is not a mass eigenstate, but mixes with
infinitely many higher Kaluza--Klein modes $\Psi_n$. A consistent 
manifestation of a super--Higgs mechanism would require a
diagonalization of this mass matrix and an identification of
the goldstino. This problem, that has not yet been addressed in
the literature, will be solved in section 4. 

This resolution of the puzzles clarifies some of the other
questions of the approach.

\begin{itemize}

\item  The nonlocality of the breakdown shows some resemblance
to the breakdown of supersymmetry via the 
Scherk--Schwarz \cite{ScherkSchwarz} mechanism. Here,
however, the real goldstino of the spontaneous breakdown of
supersymmetry can be unambiguously identified.

\item The possibility to cancel the supersymmetry breakdown
on a distant wall by a vev on the local wall tells us,
that the mass splittings of broken supersymmetry have to
be of order of the gravitino mass $m_{3/2}$ on {\it both}
walls.

\item In terms of the physical quantities there is no real
extra suppression, once we separate the walls by a large
distance $R$. In the limit $R\rightarrow\infty$
we will have $M_{\rm Planck}\rightarrow\infty$ as well.
The suppression of the soft breaking parameters will
always be gravitational, as given in (\ref{eq:gravitinomass2}).

\item In general, when we have a system of many separated 
branes with potential sources of supersymmetry breakdown,
the actual breakdown will be obtained by the sum of these
contributions. The averaging proceedure will be very
useful to decide whether supersymmetry is broken or not. The 
identification of the goldstino, however, is more
difficult and requires a careful calculation.

\item A successful implementation of the super--Higgs mechanism
will require some fields other than gravitino and graviton in 
the bulk\footnote{Usually they arise as modes of the higher
dimensional supergravity multiplet.}. 
This implies that in the absence of such fields
(as has been considered in \cite{randall}) a consistent
spontaneous breakdown of supergravity might not be achieved.

\end{itemize}

In the following sections we will show how the goldstino and
gravitino can be defined in the correct way. We shall do
the explicit calculations in the framework of the heterotic
M-theory, although a similar calculation will apply under
more general circumstances (like the inclusion of 5-branes or
the consideration of multi-D-brane systems in Type I theory), 
which we shall briefly discuss in section 5. As the source of 
supersymmetry we consider the mechanism of gaugino condensation.
Again this just should represent a generic breakdown of
supersymmetry in this specific example.

\section{Gravitino in the case of unbroken supersymmetry}

The low energy limit of the heterotic M--theory is given  
by the following lagrangian
\bea
{\cal L}&=& \frac{1}{\kappa^2}\int_{M^{11}}d^{11}x\sqrt g
\left[
-\frac{1}{2}R-\frac{1}{2}\bar\Psi_I\Gamma^{IJK}D_J\left(\frac{\Omega+
\hat\Omega}{2}\right)\Psi_K-\frac{1}{48}
{\tilde G}_{IJKL}{\tilde G}^{IJKL}\right.\nn\\
&& \quad -\frac{\sqrt 2}{384}\left(\overline\Psi_I\Gamma^{IJKLMN}\Psi_N
+12\overline\Psi^J\Gamma^{KL}\Psi^M\right)\left({\tilde G}_{JKLM}
+\hat{\tilde G}_{JKLM}\right)\nn\\
&& \quad\quad \left.{}-\frac{\sqrt 2}{3456}\epsilon^{I_1I_2\ldots I_{11}}
{\tilde C}_{I_1I_2I_3}{\tilde G}_{I_4\ldots I_7}{\tilde G}_{I_8\ldots
I_{11}}
\right]
\label{lagr}
\\
&+&
\frac{1}{4\pi(4\pi\kappa^2)^{2/3}}
\sum_{i=1}^2
\int_{M^{10}_i}d^{10}x\sqrt g
\left[
-\frac{1}{4}F^a_{iAB}F^{a\,AB}_i-\frac{1}{2}\overline\chi^a_i
\Gamma^AD_A(\hat\Omega)\chi^a_i\right.\nn\\
&& \quad \left.\qquad{}-\frac{1}{8}\overline\Psi_A\Gamma^{BC}
\Gamma^A\left(F^a_{iBC}+
\hat F^a_{iBC}\right)\chi^a_i+\frac{\sqrt 2}{48}
\left(\overline\chi^a_i\Gamma^{ABC}
\chi^a_i\right)\hat{\tilde  G}_{ABC\,11}
\right]
\,.\nn
\eea
where $I,J,K,\ldots=1,2,\ldots, 11$; $A,B,C,\ldots=1,2,\ldots, 10$; 
and $i=1,2$ counts the 10--dimensional boundaries (walls) of the space. 
The first integral describes the supergravity in the 11--dimensional bulk 
while the second one describes interactions with the super Yang--Mills 
fields living on two 10--dimensional walls. 
Our signature is $(-,+,\ldots,+)$.  
In the above lagrangian only the two first terms in the long wavelength 
expansion are kept. They are of relative order $\kappa^{2/3}$. 
All higher order terms (order $\kappa^{4/3}$ or higher) 
will be consistently dropped in this paper. 

We work in the upstairs approach in which the 11--dimensional integrals 
are defined as 
\be
\int d^{11}x = \frac{1}{2}\int_{-\pi\rho}^{\pi\rho}dx^{11}\int d^{10}x.
\ee
and we use the $Z_2$ symmetry conditions\footnote
{In the downstairs approach we would use
$$
\int d^{11}x = \int_0^{\pi\rho}dx^{11}\int d^{10}x
$$
and the appropriate boundary conditions instead of $Z_2$ symmetry.
}
:
\bea
\Psi_A(-x^{11})&=&+\Gamma^{11}\Psi_A(x^{11})
\nn\\
\Psi_{11}(-x^{11})&=&-\Gamma^{11}\Psi_{11}(x^{11})
\nn\\
{\tilde G}_{ABCD}(-x^{11})&=&-{\tilde G}_{ABCD}(x^{11})
\label{Z2}
\\
{\tilde G}_{BCD11}(-x^{11})&=&+{\tilde G}_{BCD11}(x^{11})
\nn
\eea
The Bianchi identity in this approach is modified on both walls: 
\be
d{\tilde G}_{11\,ABCD}=-\frac{3\sqrt 2}{2\pi}\left(\frac{\kappa}{4\pi}
\right)^{2/3}
\sum_{i=1}^2
\delta(x^{11}-x^{11}_i)\left(\tr\;F_{i[AB}F_{iCD]}
-\frac{1}{2}R_{[AB}R_{CD]}
\right)\,.
\label{dgeq}
\ee
where $x^{11}_i$ is a position of the $i$--th wall.  

The fields $\Psi_{\mu}$ and $\Psi_{11}$ as 
11--dimensional Majorana spinors have 32
components. Imposing $SU(3)$ invariance on Calabi--Yau reduces this
number to 8 components -- they can be assembled into two sets with 4
components each distinguished by 10-dimensional chirality. 
Some of the formulae given below are valid only after
imposing $SU(3)$ invariance but we treat all spinors as 11--dimensional along
the way. Only at the very end, after
compactification to 4 dimensions we assemble each set into a 4--dimensional
Majorana spinor to give the final formula for 
the effective 4--dimensional action for spinors.

This theory is defined in 11 dimensions (\ref{lagr}) so 7 dimensions must be 
compact. The spacetime is given (in the lowest approximation) 
by the  product: 
$M^{4} \times X^6 \times S^1/Z_2$ where $X^6$ is a Calabi--Yau manifold 
and $S^1/Z_2$ is the interval between the two walls. 
For simplicity we will use a truncation and reduction method 
\cite{trunc,NOY1,NOY2} 
instead of compatifying on a Calabi--Yau manifold. 
The 11--dimensional metric is given in this case by
\be
g^{(11)}_{MN} =
\left(
\ba{ccc}
e^{-\gamma} e^{-2\sigma} g_{\mu\nu} & & \\
 & e^\sigma g_{mn} & \\
 & & e^{2\gamma} e^{-2\sigma}
\ea
\right)
\label{g11}
\ee
At order $\kappa^{2/3}$ the moduli $\gamma$ and $\sigma$ 
depend linearly on $x^{11}$ and the 11--dimensional spacetime 
is no longer a direct product of the three factors. 

Let us now describe a process of reduction of the theory to 4 dimensions. 
We will be mainly interested in the fermion fields since our main goal 
is to identify the 4--dimensional (massless or massive, depending 
on supersymmetry breaking) gravitino. It turns out that in the resulting 
lagrangian all the fermion fields are mixed. To diagonalize that lagrangian 
one has to make a number of field redefinitions. In order to make the 
result of the calculation more transparent we will perform the 
appropriate redefinitions (necessary to get the final 
result in a diagonal form) step by step. 

Let us start with the following redefinition of the gravitino fields
\bea
\Psi_\mu &=& e^{-\gamma/4}e^{-\sigma/2}
\left(\psi_\mu + \frac{1}{\sqrt{6}}\Gamma_\mu\psi_{11}\right)\nn\\
\Psi_{11}&=&
-\frac{2}{\sqrt{6}}e^{5\gamma/4}e^{-\sigma/2}
\Gamma^{11}\psi_{11}
\,.
\label{psired1}
\eea
As a result the kinetic term of the lagrangian 
(\ref{lagr}) reads:
\bea
-\frac{1}{2}\!\!\!\!&e_{11}&\!\!\!\!\overline\Psi_I
\Gamma^{IJK}D_J\left(\frac{\Omega+
\hat\Omega}{2}\right)\Psi_K=
\nn\\
&-&\!\!\!\!\!\frac{1}{2}e_4\overline\psi_{\mu}\Gamma^{\mu\nu\rho}D_{\nu}
\psi_{\rho}
-\frac{1}{2}e_4\overline\psi_{11}\Gamma^{\mu}D_{\mu}\psi_{11}
\nn\\
&+&\!\!\!\!\!\frac12 e_4e^{-3\gamma/2}
\psi_{\mu}\Gamma^{\mu\nu}\Gamma^{11}\pa_{11}\psi_{\nu} 
-\frac{\sqrt{6}}{4} e_4e^{-3\gamma/2}
\psi_{11}\Gamma^{\mu}\Gamma^{11}\pa_{11}\psi_{\nu}
\nn\\
&-&\!\!\!\!\!\frac{\sqrt{6}}{4} e_4e^{-3\gamma/2}
\psi_{\mu}\Gamma^{\mu}\Gamma^{11}\pa_{11}\psi_{11}
+e_4e^{-3\gamma/2}
\psi_{11}\Gamma^{11}\pa_{11}\psi_{11}
\nn\\
&-&\!\!\!\!\!\frac{\sqrt{6}}{4} e_4e^{-3\gamma/2}
\psi_{\mu}\Gamma^{\nu}\Gamma^{\mu}\psi_{11} \pa_{\nu}\gamma
+\frac{\sqrt{6}}{4} e_4e^{-3\gamma/2}
\psi_{\mu}\Gamma^{\mu}\Gamma^{11}\psi_{11} \pa_{11}\gamma
\label{kinpsi}
\eea
Let us introduce 
\bea
\psi_{\mu}&=&\psi_{\mu}^{+}(x^{11})+
\psi_{\mu}^{-}(x^{11})\nn\\
\psi_{11}&=&
\psi_{11}^{-}(x^{11})+\psi_{11}^{+}(x^{11})
\label{chirpsi}
\eea
(the signs "$+$" and "$-$" denote the chirality with respect 
to $\Gamma^{11}$).
The relations (\ref{Z2}) show that
\bea
\psi_{\mu}^{+}(-x^{11})&=&+\psi_{\mu}^{+}(x^{11})\nn\\
\psi_{\mu}^{-}(-x^{11})&=&-\psi_{\mu}^{-}(x^{11})\nn\\
\psi_{11}^{+}(-x^{11})&=&-\psi_{11}^{+}(x^{11})\nn\\
\psi_{11}^{-}(-x^{11})&=&+\psi_{11}^{-}(x^{11})
\label{psichrefl}
\eea
Therefore the zero modes are possible only for $\psi_{\mu}^{+}$ and
$\psi_{11}^{-}$. All fields have an implicit
dependence on $x^{\mu}$ and we will often omit $x^{11}$ dependence of
the fields.

Let us now consider switching on vev of the tensor field $\tilde G$. 
As explained in \cite{Witten} due to the nonzero r.h.s.\ 
of eq.\ (\ref{dgeq}) fields $\tilde G_{a\overline b c\overline d}$ 
(where $a,\ldots$ are holomorphic and $\bar a, \ldots$ are 
antiholomorphic indices on $X^6$) acquire nonzero vev 
satisfying
\be
\langle\Gamma^{a\overline b c\overline d}
\tilde G_{a\overline b c\overline d}\rangle\psi=
-\frac{\alpha}{2}\epsilon(x^{11})\psi
\label{gvev}
\ee
on any $SU(3)$ invariant spinor $\psi$ (with arbitrary chirality).
The parameter $\alpha$ is given by
\be
\omega^{AB}\omega^{CD}\tilde G_{ABCD}
=4\omega^{a\overline b}\omega^{c\overline d}
\tilde G_{a\overline b c\overline d}
=\alpha
\,.
\ee
In the case of unbroken supersymmetry there is a relation between 
$\alpha$ and the slopes of $\gamma(x^{11})$ and $\sigma(x^{11})$ 
\cite{Witten,NOY2}:
\be
\pa_{11}\gamma
=-\pa_{11}\sigma
=\frac{\sqrt{2}}{24}\alpha\,\epsilon(x^{11})
\label{gamalpha}
\ee

In order to get the canonical kinetic terms for $G$ we 
make the following redefinitions:
\bea
{\tilde G}_{a\bar a b\bar b}&=& e^{\gamma/2}e^{3\sigma}\left(
G_{a\bar a b\bar b}-\frac{\sqrt{3}}{3}\overline\psi_{11}^-\Gamma^{\mu}
\Gamma_{a\bar a b\bar b}\psi_{\mu}^{-}
-\frac{2\sqrt{2}}{3}\overline\psi_{11}^+
\Gamma_{a\bar a b\bar b}\psi_{11}^{-}\right)\nn\\
{\tilde G}_{11abc}&=& e^{3\gamma/2}e^{3\sigma/2}G_{11abc}\nn\\
{\tilde G}_{11\bar a\bar b\bar c}&=&
e^{3\gamma/2}e^{3\sigma/2}G_{11\bar a\bar b\bar c} 
\label{gred}
\eea
The unusual additional terms in the redefinition of 
${\tilde G}_{a\bar a b\bar b}$ are 
necessary to cancel some of the fermion nondiagonal terms 
(as will be shown after eq.\ (\ref{gkin})).

In this paper we keep track of only (2,2,0), 
(3,0,1) and (0,3,1) components of $<G>$. It was shown by Witten 
\cite{Witten} that the
presence of the vacuum expectation value for the (2,2,0) component of $G$
does not break supersymmetry when the fuctions $\gamma$ and $\sigma$
have definite dependence on $x^{11}$ (see eq.\ (\ref{gamalpha})). 
The presence of the (3,0,1) and
(0,3,1) components of $<G>$ located on the walls generically breaks
supersymmetry 
and in the next section we will provide the explicit formula for
the mass of the gravitino, its expansion in the Kaluza--Klein modes and the
disappearance of one spin 1/2 state (super--Higgs mechanism). 
In this section we will consider the case of unbroken supersymmetry 
discussed by Ho\v{r}ava \cite{Horava}. 
We now need the couplings of $<G>$ to the fermion fields
\bea
\!\!\!&-&\!\!\!\frac{\sqrt2}{192}e_{11}
\overline\Psi_I\Gamma^{IJKLMN}\Psi_J {\tilde G}_{KLMN}=\nn\\
\!\!\!&-&\!\!\!
\frac{\sqrt{2}}{96}e_4\left(3\overline\psi_{\mu}\Gamma^{\mu\nu}
\Gamma^{ab\overline c\overline d}\psi_{\nu}+
\sqrt{6}\,\overline\psi_{\mu}\Gamma^{\mu}
\Gamma^{ab\overline c\overline d}\psi_{11}+
2\overline\psi_{11}\Gamma^{ab\overline c\overline d}
\psi_{11}\right)<G_{ab\overline c\overline d}>
\nn\\
\!\!\!&-&\!\!\!
\frac{2\sqrt{2}}{96}e_4\left(\overline\psi_{\mu}\Gamma^{\mu\nu}
\Gamma^{abc}\Gamma^{11}\psi_{\nu}+
\sqrt{6}\,\overline\psi_{\mu}\Gamma^{\mu}
\Gamma^{abc}\Gamma^{11}\psi_{11}-
2\overline\psi_{11}\Gamma^{abc}\Gamma^{11}\psi_{11}\right)
<G_{abc11}>
\nn\\
\!\!\!&-&\!\!\!
\frac{2\sqrt{2}}{96}e_4\left(\overline\psi_{\mu}\Gamma^{\mu\nu}
\Gamma^{\overline a\overline b\overline c}\Gamma^{11}\psi_{\nu}+
\sqrt{6}\,\overline\psi_{\mu}\Gamma^{\mu}
\Gamma^{\overline a\overline b\overline c}\Gamma^{11}\psi_{11}-
2\overline\psi_{11}\Gamma^{\overline a\overline b\overline c}\Gamma^{11}
\psi_{11}\right)
<G_{\overline a\overline b\overline c 11}>\nn\\
\!\!\!&+&\!\!\!\ldots
\label{psipsiG}
\eea
Using the redefinition (\ref{gred}) and the vacuum expectation value
for $G^{a\bar a b\bar b}$ (\ref{gvev}) we rewrite the kinetic term for 
$\tilde G$:
\bea
-\frac{1}{48}\!\!\!&e_{11}&\!\!\!
{\tilde G}^{IJKL}{\tilde G}_{IJKL}=\nn\\
&&-\frac{1}{12}e_4 G^{11abc}G_{11abc}
-\frac{1}{12}e_4 G^{11\bar a\bar b\bar c}G_{11\bar a\bar b\bar c}
-\frac{1}{8}e_4 G^{a\bar a b\bar b}G_{a\bar a b\bar b}\nn\\
&&-\frac{\sqrt{6}}{2}e_4
(\pa_{11}\gamma)\overline{\psi_{11}^{-}}
\Gamma^{\mu}\psi_{\mu}^{-}
- 2e_4 (\pa_{11}\gamma)\overline{\psi_{11}^{-}}\psi_{11}^{+}
+\ldots
\label{gkin}
\eea
Terms with the fermion fields in the above formula 
are necessary to cancel some of the nondiagonal terms 
coming from (\ref{kinpsi}) and (\ref{psipsiG}). 

With all these redefinitions we are now ready to evaluate the sum of
(\ref{kinpsi}), (\ref{psipsiG}) and (\ref{gkin}) 
with the vacuum expectation value for  $G_{a\overline b c\overline d}$
(\ref{gvev}) -- it is the final result for the case considered by
Witten. Using also (\ref{gamalpha}) and (\ref{gred}) we get 
the following terms bilinear in the fermionic fields
\bea
{\cal L}=&-&\!\!\!\frac12 e_4 \overline{\psi_{\mu}^{+}}\Gamma^{\mu\nu\rho}
D_{\nu}\psi_{\rho}^{+}-\frac12 e_4 \overline{\psi_{\mu}^{-}}\Gamma^{\mu\nu\rho}
D_{\nu}\psi_{\rho}^{-}\nn\\
&-&\!\!\!\frac12e_4 \overline{\psi_{11}^{-}}
\Gamma^{\mu}D_{\mu}\psi_{11}^{-}-\frac12e_4 \overline{\psi_{11}^{+}}
\Gamma^{\mu}D_{\mu}\psi_{11}^{+}\nn\\
&+&\!\!\!e_4e^{-3\gamma/2}
\psi_{\mu}^{-}\Gamma^{\mu\nu}\pa_{11}\psi_{\nu}^{+} 
-\frac{\sqrt{6}}{2} e_4e^{-3\gamma/2}
\psi_{11}^{+}\Gamma^{\mu}\pa_{11}\psi_{\nu}^{+}\nn\\
&+&\!\!\!\frac{\sqrt{6}}{2} e_4e^{-3\gamma/2}
\psi_{\mu}^{-}\Gamma^{\mu}\pa_{11}\psi_{11}^{-}
-2e_4e^{-3\gamma/2}
\psi_{11}^{+}\pa_{11}\psi_{11}^{-}
+\ldots
\label{mixeq}
\eea

Since the eleventh dimension is compact we can make the Fourier expansion
of the fields: 
\bea
\psi_{\mu}(x^{\mu},x^{11})&=&\psi_{\mu}^{0+}
+\sqrt{2}\sum_{n=1}^\infty\psi_{\mu}^{n+}\cos(n x^{11}/\rho)
+\sqrt{2}\sum_{n=1}^\infty\psi_{\mu}^{n-}\sin(n x^{11}/\rho)
\,,\nn\\
\label{psimod}\\
\psi_{11}(x^{\mu},x^{11})
&=&\psi_{11}^{0-}+\sqrt{2}\sum_{n=1}^\infty\psi_{11}^{n-}\cos(n x^{11}/\rho)
+\sqrt{2}\sum_{n=1}^\infty\psi_{11}^{n+}\sin(n x^{11}/\rho)
\,.
\nn
\eea
where the coefficients of the expansion depend only on $x^{\mu}$ (so
they correspond to 4--dimensional spinor fields).

Substituting this into eq.\ (\ref{mixeq}) and integrating over $x^{11}$  
we can see that $\psi_{\mu}^{0+}$ and $\psi_{11}^{0-}$ are massless 
-- therefore they correspond to the 4--dimensional gravitino and the massless 
spin 1/2 fermion fields. The remaining fields form the usual infinite tower 
of KK modes with masses equal to $n/\rho$. This is very similar to the 
standard Kaluza--Klein reduction. The only difference is that we had to 
redefine the tensor field $\tilde G$ (\ref{gred}) in order to remove some 
nondiagonal kinetic terms from the lagrangian. 

Let us now include a nonvanishing vev of $G_{abc11}$ and 
$G_{\overline a \overline b \overline c 11}$ fields. 
Such vevs can be generated for example by condensation of gaugino
fields living on the walls. In such a case the vev of $G_{abc11}$ on
the wall at $x^{11}=0$ is given by:
\be
\left< G_{abc11}\right>
=\frac{\sqrt{2}}{16\pi}\left(\frac{\kappa}{4\pi}\right)^{2/3}
\delta\left(x^{11}\right)\langle\chi\Gamma_{abc}\chi\rangle
\ee
An analogous formula holds for $G_{\bar a \bar b \bar c 11}$ and 
other walls.

In the case of two walls the most general formula reads:
\be
\left<G_{abc11}\right>=\epsilon_{abc}\left[G_{+}\left(\delta(x^{11})+
\delta(x^{11}-\pi\rho)\right)+
G_{-}\left(\delta(x^{11})-\delta(x^{11}-\pi\rho)\right)\right]
\label{gabcvev}
\ee 

Ho\v{r}ava \cite{Horava} discussed the case when 
the supersymmetry remains unbroken i.e. when $G_+=0$.
Let us try to see one manifestation of the unbroken supersymmetry i.e.
the massless gravitino in terms of the usual KK modes.   
Using the Fourier expansion (\ref{psimod}) and putting the nonzero 
vev $G_-$ into eqs.\ (\ref{psipsiG}) and (\ref{gkin}) we get terms which mix 
different modes
\bea
G_-
&&\!\!\!\!\!\!\!\!\!\!\left[-\overline{\psi^{0+}_\mu}
\sum_{k=1}\left(
\frac{1}{2}\Gamma^{\mu\nu}\Gamma_Y\psi^{(2k-1)+}_\nu 
+ \frac{\sqrt{6}}{4}\Gamma^\mu\Gamma_Y \psi_{11}^{(2k-1)-}
\right)\right.
\nn\\
&&\!\!\!\!\!\!\left.
+\overline{\psi_{11}^{0-}}
\sum_{k=1}\left(\frac{\sqrt{6}}{4}\Gamma^\mu\Gamma_Y\psi^{(2k-1)+}_\mu
+\Gamma_Y\psi_{11}^{(2k-1)-}\right)
\right]
+\ldots
\eea
where 
\be
\Gamma_Y=\frac{1}{6}
\left(\epsilon_{abc}\Gamma^{abc}
+\epsilon_{\bar a \bar b \bar c}\Gamma^{\bar a \bar b \bar c}\right)\,.
\ee
and $\ldots$ stand for the (quite complicated and nondiagonal 
spin 3/2 and spin 1/2) 
mass terms involving only the nonconstant KK modes. One can see that 
the constant modes $\psi_{\mu}^{0+}$ and $\psi_{11}^{0-}$ would be no 
longer massless after compactification to 4 dimensions. 
But the supersymmetry is unbroken and there 
should be the massless gravitino in the spectrum. 
One can obtain the fermions with definite masses 
by diagonalization of the mass terms. 
This, however, requires a very tedious calculation.

It turns out that it is possible to identify the massless 
states in a much simpler way. 
Before calculating the mass terms for fermions 
we perform a rotation of the fields  $\psi_{\mu}$
and $\psi_{11}$:
\bea
\psi_{\mu}&\to&
\left(1+f(x^{11})\Gamma_Y
\right)\psi_{\mu}\nn\\
\psi_{11}&\to&
\left(1-f(x^{11})\Gamma_Y
\right)\psi_{11}
\label{rotsup}
\eea
where
\be
f(x^{11})=-\frac{\sqrt{2}}{8}G_{-}\epsilon(x^{11})
\ee
Such a rotation makes the new zero mode a combination of the old 
zero mode and infinitely many excited KK modes. 
Evaluating the effect of
(\ref{gabcvev}) on (\ref{psipsiG}) and taking into account the redefinition 
(\ref{rotsup}) we recover the lagrangian (\ref{mixeq}) but now in
terms of the new fields. Hence the lagrangian (\ref{mixeq}) is the final
result in the case considered by Ho\v{r}ava (but the fields are 
those obtained after
the rotation (\ref{rotsup})). Therefore, the zero modes of the rotated
fields  $\psi_{\mu}^{+}$ and $\psi_{11}^{-}$ are now the massless
gravitino and the massless spin 1/2 fields.

It is easy now to find an explicit form of the massless fermions 
in terms of the old KK modes 
\bea
\psi^{gravitino}_\mu
&=& 
\psi_\mu^{0+} 
- \sum_{k=1}^\infty\frac{G_-}{2(2k-1)\pi}\Gamma_Y\psi_\mu^{(2k-1)-}
\nn\\
\label{supferred}\\
\psi^{(1/2)}
&=&
\psi_{11}^{0-} 
+ \sum_{k=1}^\infty\frac{G_-}{2(2k-1)\pi}\Gamma_Y\psi_{11}^{(2k-1)+}\,.
\nn
\eea
These fields are different from the constant modes which were massless 
in the case with vanishing $G_-$. This change of the massless fermions 
reflects the fact that the unbroken supersymmetry changes\footnote
{
The unbroken supersymmetry corresponds to the spinor parameter $\eta$ 
which depends on the compact coordinates. That dependence changes with 
the value of $G_-$.
} 
when we change the value of $G_-$.

\section{The super--Higgs mechanism}

In order to break the supersymmetry in this scenario we have to assume
that the condensates on opposite walls do not cancel ($G_+\ne 0$ in
the formula (\ref{gabcvev})).
In this case the fermion mass matrix is even more complicated than 
in the case with only $G_-$ nonzero. The procedure of diagonalization 
is very tedious but it turns out that, as before, it is much simpler to work 
in the 5--dimensional language. We can rotate the fields in the similar 
way as in (\ref{rotsup}):
\bea
\psi_{\mu}&\to&
\left(1+f(x^{11})\Gamma_Y
\right)\psi_{\mu}\nn\\
\psi_{11}&\to&
\left(1-f(x^{11})\Gamma_Y
\right)\psi_{11}
\label{rotnsup}
\eea
but now with
\be
f(x^{11})=-\frac{\sqrt{2}}{8}(G_{-}+G_{+})\epsilon(x^{11})+
\frac{\sqrt{2}}{4\pi\rho}G_{+}\ x^{11}
\label{fnsup}
\ee
Evaluating all the terms after the rotation (\ref{rotnsup}) we get
\bea
{\cal L}=&-&\!\!\!\frac12 e_4 \overline{\psi_{\mu}^{+}}\Gamma^{\mu\nu\rho}
D_{\nu}\psi_{\rho}^{+}-\frac12 e_4 
\overline{\psi_{\mu}^{-}}\Gamma^{\mu\nu\rho}
D_{\nu}\psi_{\rho}^{-}\nn\\
&-&\!\!\!\frac12e_4 \overline{\psi_{11}^{-}}
\Gamma^{\mu}D_{\mu}\psi_{11}^{-}-\frac12e_4 \overline{\psi_{11}^{+}}
\Gamma^{\mu}D_{\mu}\psi_{11}^{+}\nn\\
&+&\!\!\!e_4e^{-3\gamma/2}
\overline{\psi_{\mu}^{-}}\Gamma^{\mu\nu}\pa_{11}\psi_{\nu}^{+} 
-\frac{\sqrt{6}}{2} e_4e^{-3\gamma/2}
\overline{\psi_{11}^{+}}\Gamma^{\mu}\pa_{11}\psi_{\nu}^{+}\nn\\
&+&\!\!\!\frac{\sqrt{6}}{2} e_4e^{-3\gamma/2}
\overline{\psi_{\mu}^{-}}\Gamma^{\mu}\pa_{11}\psi_{11}^{-}
-2e_4e^{-3\gamma/2}
\overline{\psi_{11}^{+}}\pa_{11}\psi_{11}^{-}\nn\\
&-&\!\!\!\frac{m}2 
e_4 \overline{\psi_{\mu}^{+}}\Gamma^{\mu\nu}\Gamma_Y\psi_{\nu}^{+} 
+\frac{m}2 e_4 \overline{\psi_{\mu}^{-}}\Gamma^{\mu\nu}\Gamma_Y\psi_{\nu}^{-}
-\frac{m\sqrt{6}}{2} e_4
\overline{\psi_{\mu}^{-}}\Gamma^{\mu}\Gamma_Y\psi_{11}^{+}\nn\\
&-&\!\!\!\frac{m\sqrt{6}}{2} e_4
\overline{\psi_{\mu}^{+}}\Gamma^{\mu}\Gamma_Y\psi_{11}^{-}
+m\ e_4\overline{\psi_{11}^{+}}\Gamma_Y\psi_{11}^{+}
-m\ e_4\overline{\psi_{11}^{-}}\Gamma_Y\psi_{11}^{-}+\ldots
\label{nsupl}
\eea
where
\be
m=\frac{\sqrt{2}}{4\pi\rho}G_{+}
\label{mass}
\ee
is the mass of the lightest spin 3/2 state -- the gravitino. 
Zero modes of the rotated fields are now the lowest--lying states and 
from the rotation (\ref{rotnsup}) 
we can read off their composition in terms of the standard KK modes:
\bea
\psi^{gravitino}_\mu
&=& 
\psi_\mu^{0+} 
- \sum_{k=1}^\infty\frac{G_{-}+G_{+}}{4\pi k}
\left[1-(-1)^k\right]\Gamma_Y\psi_\mu^{k-}
+\sum_{k=1}^\infty\frac{G_{+}}{2\pi k}(-1)^{k+1}\psi_\mu^{k+}
\nn\\
\label{nsupferred}\\
\psi_{11}^{goldstino}
&=&
\psi_{11}^{0-} 
+ \sum_{k=1}^\infty\frac{G_{-}+G_{+}}{4\pi k}
\left[1-(-1)^k\right]\Gamma_Y\psi_{11}^{k+}
-\sum_{k=1}^\infty\frac{G_{+}}{2\pi k}(-1)^{k+1}\psi_{11}^{k-}
\nn
\eea

A remark is in order here: the formula for the gravitino mass 
(\ref{mass}) is the same as in the naive approach (just taking the
zero mode and not performing the rotation (\ref{rotnsup})).
This mass is already a $\kappa^{2/3}$ effect, so corrections 
of the next order in $\kappa^{2/3}$ (or inversely proportional 
to $M_{11}$) must be dropped. 
In general we could expect other corrections 
which should be kept e.g.\ proportional to $\pi\rho\gg M_{11}^{-1}$. 
However, it turns out that no such corrections come  
from the proper diagonalization of the mass matrix. 
On the other hand, we find corrections to the composition of the 
mass eigenstates 
(in perturbation theory corrections to the eigenfunctions are ususally more
difficult to obtain  than corrections to the
eigenvalues).

Let us now discuss the super--Higgs effect which should take place 
when supersymmetry is spontaneously broken as in the present case. 
In the Lagrangian (\ref{nsupl}) there are many terms containing 
both the gravitino and the goldstino fields. 
There should exist a way of ``eating'' the massless goldstino 
and leaving only the massive gravitino. 
And indeed let us define 
\be
\psi^{(3/2)}_\mu=
\psi_{\mu}^{gravitino}+\frac{2}{\sqrt{6}m}\Gamma_YD_\mu\psi_{11}^{goldstino}
+\frac{1}{\sqrt{6}}\Gamma_{\mu}\psi_{11}^{goldstino}
\,.
\ee
Then as a part of (\ref{nsupl}) we obtain a lagrangian for the 
massive gravitino
\be 
-\frac12 e_4 \overline{\psi^{(3/2)}_\mu}\Gamma^{\mu\nu\rho}
D_{\nu}\psi^{(3/2)}_\rho-
\frac{m}2 e_4 \overline{\psi^{(3/2)}_\mu}\Gamma^{\mu\nu}
\psi^{(3/2)}_\nu
\label{masl}
\ee
and the field $\psi_{11}^{goldstino}$ completely 
disappears from (\ref{nsupl}). This is precisely the super--Higgs
effect since the goldstino provided the degrees of freedom needed for
the massless gravitino to become massive.

\section{Discussion}
In the case with the gaugino condensate present only at the hidden wall 
the gravitino mass is given by 
\be
m_{3/2}=\frac{1}{16(4\pi)^{(5/3)}}\frac{\Lambda^3}{M_{Pl}^2}
< 10^{-3}\frac{\Lambda^3}{M_{Pl}^2}
\ee
We need the scale of the condensate $\Lambda$ to be of the order 
$10^{14}$ GeV to get the gravitino mass of about 1 TeV. 
If we assume that $\Lambda$ is not bigger than the GUT scale 
($10^{16}$ GeV) then we obtain the upper bound on the gravitino mass 
of the order of $10^5$ TeV. A value of $\Lambda$ much larger than the
GUT scale (which is comparable to the 11-dimensional Planck scale $M_{11}$)
will not be meaningful in this framework.

In the above we have shown how to identify the gravitino and goldstino 
fields in the case with arbitrary gaugino condensates 
at two 10--dimensional walls of the 11--dimensional spacetime. 
This procedure can be easily generalized to more complicated situations.
We can analyse, for example, a model in which supersymmetry breaking 
sources are present not only at the walls but also at some branes 
located along the eleventh dimension interval. In this case we have 
to perform a field redefinition similar to the rotation 
(\ref{rotnsup}) 
but with modified function $f(x^{11})$ and/or matrix $\Gamma_Y$. 
Those modifications should be chosen in such a way that the terms 
containing $\partial_{11}$ present in the lagrangian 
(\ref{mixeq}) 
exactly cancel (locally) all $\delta$--like sources of supersymmetry 
breaking. An additional constraint on the function $f$ comes from the 
$Z_2$ symmetry and relates its values at $x^{11}=-\pi\rho$ and 
$x^{11}=+\pi\rho$. To fulfill such a constraint we generally have 
to add a linear part to $f$ (like in (\ref{fnsup})). The compactification 
to 4 dimensions is quite straightforward in terms of the rotated 
fields because the KK modes of those fields are the mass 
eigenstates.
The linear term in $f$ gives rise 
to the gravitino mass after compactification to 4 dimensions. 
This mass is zero (and supersymmetry is unbroken) only if all 
the sources add up to zero. 

Let us now compare the above discussed mechanism of supersymmetry 
breaking to the Scherk--Schwarz mechanism \cite{ScherkSchwarz}. 
There is one similarity: in both cases the 4--dimensional fields 
(for example the gravitino) are obtained from higher dimensional 
fields with nontrivial dependence on the compact coordinate(s). 
But the origin of this dependence is very different. In the 
Scherk--Schwarz mechanism we just assume some specific dependence or,
in other words, we keep only one (nonconstant) KK mode and drop 
all the other KK modes (also the constant one). 
The mass of the gravitino is 
equal to this KK mass and as a result supersymmetry is explicitly 
broken. In this paper, on the contrary, 
we keep all the KK modes. They mix due to the supersymmetry 
breaking sources (like a vev of $G$) and we identify the gravitino 
as the lightest mass eigenstate with spin 3/2. Supersymmetry is 
broken spontaneously and the goldstino is ``eaten'' by the 
super--Higgs mechanism as was shown explicitly in the previous 
section. It is thus 
possible to take into account effects of other (heavier) spin 3/2 
states. The mechanism is motivated by the
dynamics of the higher dimensional theory. Modified Bianchi identities
and the perfect square structure in the lagrangian provide
justification for the nontrivial background of the $G$ field. In this
background we are able to perform explicit calculations identifying
the lowest--lying gravitino and all the heavier states.

\newpage
\noindent
{\Large \bf Acknowledgements}
\vskip .5cm
H.P.N. was partially supported by the European Commission 
TMR-programs ERBFMRX-CT96-0045 and CT96-0090. 
K.A.M. was partially supported by the Polish KBN grant 2P03B 03715
(1998-2000).
M.O. was partially supported by the Polish grant 
KBN 2 P03B 052 16 (1999-2000).

\end{document}